\documentclass[final,5p,times,twocolumn]{elsarticle}
\usepackage{lineno,hyperref,amsmath,mathtools,amsthm,amssymb,amsfonts,ragged2e,color,subfig}
\journal{``Contributions to Plasma Physics"}
\begin{document}
\begin{frontmatter}
\title{Magnetized ion-acoustic shock waves in degenerate quantum plasma}
\author{S. Jahan$^{1,*}$, T.S. Roy$^{2,**}$, B.E. Sharmin$^{1,***}$, N.A. Chowdhury$^{3,\dag}$, A. Mannan$^{1,\ddag}$, and A.A. Mamun$^{1,\S}$}
\address{$^1$ Department of Physics, Jahangirnagar University, Savar, Dhaka-1342, Bangladesh\\
$^{2}$Department of Physics, Bangladesh University of Textiles, Tejgaon Industrial Area, Dhaka-1000, Bangladesh\\
$^{3}$Plasma Physics Division, Atomic Energy Centre, Dhaka-1000, Bangladesh\\
e-mail: $^{*}$jahan88phy@gmail.com, $^{**}$tanu.jabi@gmail.com, $^{***}$sharmin114phy@gmail.com,\\
$^{\dag}$nurealam1743phy@gmail.com, $^{\ddag}$abdulmannan@juniv.edu, $^{\S}$mamun\_phys@juniv.edu}
\begin{abstract}
A theoretical investigation has been carried out to examine the ion-acoustic shock waves (IASHWs) in a
magnetized degenerate quantum plasma system containing  inertialess ultra-relativistically degenerate
electrons, and inertial non-relativistic positively charged heavy and light ions. The Burgers' equation is
derived by employing reductive perturbation method. It can be seen that under consideration of
non-relativistic positively charged heavy and light ions, the plasma model supports only positive
electrostatic shock structure. It is also observed that the charge state and number density of the
non-relativistic heavy and light ions enhance the amplitude of IASHWs, and the steepness
of the shock profile is decreased with ion kinematic viscosity ($\eta$). The findings of our present
investigation will be helpful in understanding the nonlinear propagation of IASHWs in white dwarfs and neutron stars.
\end{abstract}
\end{frontmatter}
\section{Introduction}
\label{2sec:Introduction}
The research regarding the propagation of nonlinear electrostatic excitations in degenerate
quantum plasma system (DQPS) has received a substantial attention to the plasma physicist due to its ubiquitous
existence in  white dwarfs \cite{Chandrasekhar1931a,Chandrasekhar1931b,Horn1991} and neutron
stars \cite{Chandrasekhar1931a,Chandrasekhar1931b,Horn1991}. It is believed that the components of the
DQPS  are electrons,  positively charged heavy ions (e.g., ${}^{56}_{26}{\mbox{Fe}}$ \cite{Vanderburg2015},
${}^{85}_{37}{\mbox{Rb}}$ \cite{Witze2014}, ${}^{96}_{42}{\mbox{Mo}}$ \cite{Witze2014}), and
positively charged light ions (e.g., ${}^{1}_{1}{\mbox{H}}$ \cite{Fletcher2006,Killian2006}, ${}^{4}_{2}{\mbox{He}}$ \cite{Flower1994},
 ${}^{12}_{6}{\mbox{C}}$ \cite{Koester1990,Koester2002}). A number of authors investigated nonlinear
waves in DQPS having positively charged heavy and light ions, and electrons \cite{Mamun2018,Mamun2019,Islam2017,Islam2017a}.

The characteristics of DQPS are comprehensively governed by the number density of the plasma species of DQPS,
and it has been observed that the electron number density in white dwarfs is in the order of
$10^{30}$ $\mbox{cm}^{-3}$ to $10^{39}$ $\mbox{cm}^{-3}$, and even more in neutron stars \cite{Koester1990,Koester2002}.
The dynamics of these high-dense plasma species in DQPS can be predicted by the Heisenberg uncertainty principle
and Pauli exclusion principle, and under consideration of these two principles, the plasma species can create degenerate pressure which
is readily outward directional, and is not similar to the thermal pressure in normal plasmas. In extremely high-dense plasma, the degenerate pressure
usually exceeds the thermal pressure. Therefore, the degenerate pressure has to be taken into account
to model the dynamics of the DQPS. The degenerate pressure associated with degenerate electrons, heavy ions, and
light ions can be given by \cite{Chandrasekhar1931b}
\begin{eqnarray}
&&\hspace*{-1.3cm}P_s=\acute{K_s}N_s^\gamma,
\label{eq:C4:01}\
\end{eqnarray}
where $s$ represents electron or heavy ion or light ion species, i.e., $s=e$
for the electron species, $s=1$ for the heavy ion species, and $s=2$ for
the light ion species;
\begin{eqnarray}
&&\hspace*{-1.3cm}\gamma=\frac{5}{3};~~~\acute{K_s}=\frac{3}{5}\Big(\frac{\pi}{3}\Big)^\frac{1}{3}\frac{\pi\hbar^2}{m_s}\simeq\frac{3}{5}\Lambda_{cs}\hbar c,
\label{eq:C4:02}\
\end{eqnarray}
for non-relativistic limit [with $\Lambda_{cs}=\pi\hbar/m_sc$, $\hbar$ is the Planck constant ($h$) divided by $2\pi$,
$m_s$ is the mass of species $s$, and c is the speed of light in vacuum], and
\begin{eqnarray}
&&\hspace*{-1.3cm}\gamma=\frac{4}{3};~~\acute{K_s}=\frac{3}{4}\Big(\frac{\pi^2}{9}\Big)^\frac{1}{3}\hbar c\simeq\frac{3}{4}\hbar c,
\label{eq:C4:03}\
\end{eqnarray}
for ultra-relativistic limit \cite{Islam2017,Islam2017a}. The degenerate pressure
depends only on the number density of the plasma species but not on their temperature \cite{Islam2017,Islam2017a}.
For stable configuration of the DQPS, the outward directional
degenerate pressure is counter-balanced  by the inward gravitational pressure.

The electrostatic shock wave profile, which may arise due to the Landau damping and
kinematic viscosity of the medium, is governed by the Burgers' equation \cite{Atteya2017,Abdelwahed2016,Tantawy2016,Dev2018}.
Atteya \textit{et al.} \cite{Atteya2017} examined the ion-acoustic (IA) shock waves (IASHWs) in DQPS, and
reported that the amplitude of the positive shock profile increases with the increase of electron number density.
Abdelwahed \textit{et al.} \cite{Abdelwahed2016} investigated IASHWs in non-thermal plasma, and found that
the steepness of the shock profile decreases with ion kinematic viscosity.

The external magnetic field has been considered to investigate the electrostatic
shock \cite{Bains2014,Bains2015,Hossen2017} and solitary \cite{Shaukat2017,Ashraf2013} waves in plasmas.
Hossen \textit{et al.} \cite{Hossen2017} examined the IASHWs in the presence of
external magnetic field, and highlighted that the amplitude of IASHWs increases with
increasing the angle between the wave propagation vector and the direction of external
magnetic field (via $\delta$). Shaukat \cite{Shaukat2017} studied IA solitary waves in degenerate
magneto-plasma. Ashraf \textit{et al.} \cite{Ashraf2013} observed that the amplitude
of the electrostatic shock wave increases with oblique angle.

Recently, Islam \textit{et al.} \cite{Islam2017a} investigated envelope solitions in a three-component
DQPS containing relativistically degenerate electrons, positively charged heavy and light ions.
To the best knowledge of the authors, no attempt has been made to study IASHWs in a magnetized
DQPS having positively charged non-relativistic heavy and light ions, and ultra-relativistically
degenerate electrons. Therefore, the aim of our present investigation is to derive the Burgers'
equation and by employing its shock solution, we will numerically analyze the IASHWs in a magnetized DQPS.

The manuscript is organized in the following way: The governing equations are described in section \ref{2sec:Governing Equations}.
The derivation of the Burgers' equation and its shock solution are demonstrated in section \ref{2sec:Derivation of Burgers equation}.
The results and discussion  are presented  in section \ref{2sec:Results and Discussions}. The  conclusion is provided in section \ref{2sec:Conclusion}.
\section{Governing Equations}
\label{2sec:Governing Equations}
We consider a magnetized DQPS consisting of inertial positively charged non-relativistic
heavy ions (mass $m_1$; charge $q_1=+eZ_1$; number density $N_1$; pressure $P_1$), positively charged
non-relativistic light ions (mass $m_2$; charge $q_2=+eZ_2$; number density $N_2$; pressure $P_2$), and inertialess
ultra-relativistically degenerate electrons (mass $m_e$; charge $-e$; number density $N_e$; pressure $P_e$);
where $Z_1$ ($Z_2$) is the charge state of the heavy (light) ion.
We also assume an uniform external magnetic field $\mathbf{B}$ in the direction of $z$-axis
($\mathbf{B}={B_0}\hat{z}$). The propagation of IASHWs is governed  by the following equations:
\begin{eqnarray}
&&\hspace*{-1.3cm}\frac{\partial N_{1}}{\partial T}+\tilde\nabla \cdot(N_{1}{U}_{1})=0,
\label{eq:C4:04}\\
&&\hspace*{-1.3cm}\frac{\partial U_{1}}{\partial T}+({U}_{1}\cdot\tilde\nabla){U}_{1}=-\frac{Z_{1}e}{m_{1}}\tilde\nabla\tilde{\Phi} +\frac{Z_{1}eB_0}{m_{1}}({U}_{1}\times\hat{z})
\nonumber\\
&&\hspace*{1.7cm}-\frac{1}{m_{1}N_{1}}\tilde{\nabla}P_{1}+\tilde\eta\tilde\nabla^2U_1,
\label{eq:C4:05}\\
&&\hspace*{-1.3cm}\frac{\partial N_{2}}{\partial T}+\tilde\nabla \cdot(N_{2}{U}_{2})=0,
\label{eq:C4:06}\\
&&\hspace*{-1.3cm}\frac{\partial {U}_{2}}{\partial T}+({U}_{2}\cdot\tilde\nabla){U}_{2}=-\frac{Z_{2}e}{m_{2}}\tilde\nabla\tilde{\Phi} +\frac{Z_{2}eB_0}{m_{2}}({U}_{2}\times\hat{z})
\nonumber\\
&&\hspace*{1.52cm}-\frac{1}{m_{2}N_{2}}\tilde{\nabla}P_{2}+\tilde\eta\tilde\nabla^2U_2,
\label{eq:C4:07}\\
&&\hspace*{-1.3cm}\tilde\nabla^2\tilde\Phi=4\pi e(N_e-Z_2N_2-Z_1N_1),
\label{eq:C4:08}\
\end{eqnarray}
where ${U}_1$ (${U}_2$) is the fluid speed of heavy (light) ion; $\tilde\Phi$ is the electrostatic wave potential;
and $\tilde\eta$ is the kinematic viscosity for heavy and light ions, and for simplicity, we have assumed
$\tilde\eta\simeq\tilde\eta_1/m_1N_1\simeq\tilde\eta_2/m_2N_2$. The equation for the degenerate
electron can be expressed as
\begin{eqnarray}
&&\hspace*{-1.3cm}\tilde\nabla\tilde\Phi-\frac{1}{eN_e}\tilde\nabla P_e=0.
\label{eq:C4:09}\
\end{eqnarray}
Now, we have introduced the normalizing parameters as follows: $n_1\rightarrow N_1/n_{10}$; $n_2\rightarrow N_2/n_{20}$; $n_e\rightarrow N_e/n_{e0}$; $u_1\rightarrow U_1/C_1$; $u_2\rightarrow U_2/C_1$; $\phi\rightarrow e\tilde\Phi/m_ec^2$; $t\rightarrow T/\omega_{p1}^{-1}$; $\nabla\rightarrow\tilde\nabla/\lambda_{D1}$; $\eta=\tilde\eta/\omega_{p1}\lambda_{D1}^2$ \big[where $C_1=(Z_1m_ec^2/m_1)^{1/2}$; the plasma frequency $\omega_{p1}^{-1}=(m_1/4\pi Z_1^2e^2n_{10})^{1/2}$; the Debye length $\lambda_{D1}=(m_ec^2/4\pi Z_1e^2n_{10})^{1/2}$\big]. At equilibrium, the
 quasi-neutrality condition can be written as $n_{e0}\simeq Z_1n_{10}+Z_2n_{20}$. By using these normalizing parameters, Eqs. \eqref{eq:C4:04}-\eqref{eq:C4:08} can be expressed as
\begin{eqnarray}
&&\hspace*{-1.3cm}\frac{\partial n_{1}}{\partial t}+\nabla \cdot(n_{1}u_{1})=0,
\label{eq:C4:10}\\
&&\hspace*{-1.3cm}\frac{\partial}{\partial t}(u_{1})+(u_{1}\cdot\nabla)u_{1}=-\nabla\phi +\Omega_{c1}(u_{1}\times\hat{z})
\nonumber\\
&&\hspace*{1.8cm}-\frac{\mu_1\acute{K_1}}{n_{1}}\nabla n_{1}^\alpha+\eta\nabla^2{u_{1}},
\label{eq:C4:11}\\
&&\hspace*{-1.3cm}\frac{\partial n_{2}}{\partial t}+\nabla \cdot(n_{2}u_{2})=0,
\label{eq:C4:12}\\
&&\hspace*{-1.3cm}\frac{\partial}{\partial t}(u_{2})+(u_{2}\cdot\nabla)u_{2}=-\mu_2\nabla\phi +\mu_2\Omega_{c1}(u_{2}\times\hat{z})
\nonumber\\
&&\hspace*{1.6cm}-\frac{\mu_1 \acute{K_2}}{n_{2}}\nabla n_{2}^\alpha+\eta\nabla^2{u_{2}},
\label{eq:C4:13}\\
&&\hspace*{-1.3cm}\nabla^2\phi=(1+\mu_4)n_e-\mu_4n_2-n_1,
\label{eq:C4:14}\
\end{eqnarray}
where the plasma parameters are: $\Omega_{c1}=\omega_{c1}/\omega_{p1}$ [where $\omega_{c1}=Z_1eB_0/m_1$]; $\mu_1=m_1/Z_1 m_e$; $\mu_2=Z_2m_1/Z_1 m_2$; $\mu_3=n_{e0}/Z_1n_{10}$; $\mu_4=Z_2n_{20}/Z_{1}n_{10}$; $K_1={n_{10}}^{\alpha-1}\acute{K_1}/m_1c^2$; $K_2={n_{20}}^{\alpha-1}\acute{K_2}/m_2 c^2$ and $\gamma=\alpha=5/3$ (for non-relativistic limit). Now, by normalizing and integrating Eq \eqref{eq:C4:09}, the number density of the inertialess electrons can be obtained in terms of electrostatic potential $\phi$ as
\begin{eqnarray}
&&\hspace{-1.3cm}n_e=\bigg[1+\frac{\gamma_e-1}{K_3\gamma_e}\phi\bigg]^{\frac{1}{\gamma_e-1}},
\label{eq:C4:15}\
\end{eqnarray}
where $K_3={n_{e0}}^{\gamma_e-1}\acute{K_e}/m_ec^2$ and $\gamma=\gamma_e=4/3$ (for ultra-relativistic limit).
Now, expanding the right hand side of Eq. \eqref{eq:C4:15} and
substituting in Eq. \eqref{eq:C4:14}, we can write
\begin{eqnarray}
&&\hspace*{-1.3cm}\nabla^2\phi+n_1+\mu_4n_2=1+\mu_4+\sigma_1\phi+\sigma_2\phi^2+\sigma_3\phi^3+\cdot\cdot\cdot,
\label{eq:C4:16}\
\end{eqnarray}
where $\sigma_1$=$[(\mu_4+1)/\alpha K_3]$, $\sigma_2$=$[\{(\mu_4+1)(2-\gamma_e)\}/2(\alpha K_3)^2]$, and
$\sigma_3$=$[\{(\mu_4+1)(2-\gamma_e)(3-2\gamma_e)\}/6(\alpha K_3)^3]$.
\section{Derivation of the Burgers' Equation}
\label{2sec:Derivation of Burgers equation}
To study IASHWs, we derive the Burgers' equation by introducing the stretched coordinates for
independent variables as \cite{Hossen2017,Washimi1966}
\begin{eqnarray}
&&\hspace*{-1.3cm} \xi=\epsilon(l_xx+l_yy+l_zz-v_pt),
\label{eq:C4:17}\\
&&\hspace*{-1.3cm} \tau=\epsilon^2t,
\label{eq:C4:18}\
\end{eqnarray}
where $v_p$ is the phase speed and $\epsilon$ is a smallness parameter measuring the weakness of
the dissipation ($0<\epsilon<1$). The $l_x$, $l_y$, and $l_z$ (i.e., $l_x^2+l_y^2+l_z^2=1$) are
the directional cosines of the wave vector $k$ along $x$, $y$, and $z$-axes, respectively. The 
dependent variables can be expressed in power series of $\epsilon$ as \cite{Hossen2017}
\begin{eqnarray}
&&\hspace*{-1.3cm} n_1=1+\epsilon n_1^{(1)}+\epsilon^2n_1^{(2)}+\cdot\cdot\cdot,
\label{eq:C4:19}\\
&&\hspace*{-1.3cm} n_2=1+\epsilon n_2^{(1)}+\epsilon^2n_2^{(2)}+\cdot\cdot\cdot,
\label{eq:C4:20}\\
&&\hspace*{-1.3cm} u_{1x,y}=\epsilon^2 u_{1x,y}^{(1)}+\epsilon^3 u_{1x,y}^{(2)}+\cdot\cdot\cdot,
\label{eq:C4:21}\\
&&\hspace*{-1.3cm} u_{2x,y}=\epsilon^2 u_{2x,y}^{(1)}+\epsilon^3 u_{2x,y}^{(2)}+\cdot\cdot\cdot,
\label{eq:C4:22}\\
&&\hspace*{-1.3cm} u_{1z}=\epsilon u_{1z}^{(1)}+\epsilon^2 u_{1z}^{(2)}+\cdot\cdot\cdot,
\label{eq:C4:23}\\
&&\hspace*{-1.3cm} u_{2z}=\epsilon u_{2z}^{(1)}+\epsilon^2 u_{2z}^{(2)}+\cdot\cdot\cdot,
\label{eq:C4:24}\\
&&\hspace*{-1.3cm}\phi=\epsilon \phi^{(1)}+\epsilon^2 \phi^{(2)}+\cdot\cdot\cdot.
\label{eq:C4:25}\
\end{eqnarray}
Now, by substituting Eqs. \eqref{eq:C4:17}-\eqref{eq:C4:25} in Eqs. \eqref{eq:C4:10}-\eqref{eq:C4:13} and \eqref{eq:C4:16},
and collecting the terms containing $\epsilon$, the first-order equations reduce to
\begin{eqnarray}
&&\hspace*{-1.3cm}n_1^{(1)} = \frac{ l_z^2}{\big(v_p^2 - \alpha\mu_1 l_z^2 K_1\big)}\phi^{(1)},
\label{eq:C4:26}\\
&&\hspace*{-1.3cm}u_{1z}^{(1)} = \frac{v_p l_z}{\big(v_p^2 - \alpha\mu_1 l_z^2 K_1\big)}\phi^{(1)},
\label{eq:C4:27}\\
&&\hspace*{-1.3cm}n_2^{(1)} = \frac{\mu_2 l_z^2}{\big(v_p^2 - \alpha\mu_1 l_z^2 K_2\big)}\phi^{(1)},
\label{eq:C4:28}\\
&&\hspace*{-1.3cm}u_{2z}^{(1)} = \frac{\mu_2 v_p l_z}{\big(v_p^2 - \alpha\mu_1 l_z^2 K_2\big)}\phi^{(1)}.
\label{eq:C4:29}\
\end{eqnarray}
Now, the phase speed of IASHWs can be written as
\begin{eqnarray}
&&\hspace*{-1.3cm}v_p=v_{p+}= l_z \sqrt{\frac{m_2 + \sqrt{m_2^2 -4 m_1 m_3}}{2 m_1}},
\label{eq:C4:30}\\
&&\hspace*{-1.3cm}v_p = v_{p-}= l_z \sqrt{\frac{m_2 - \sqrt{m_2^2 -4 m_1 m_3}}{2 m_1}},
\label{eq:C4:31}\
\end{eqnarray}
where $m_1=\sigma_1$, $m_2=1+\mu_2 \mu_4-\alpha \sigma_1 \mu_1 K_2-\alpha \sigma_1 \mu_1 K_1$,
and $m_3=\alpha \mu_1 K_2+\alpha\mu_1 \mu_2 \mu_4K_1+\sigma_1 \alpha^2 \mu_1^2K_1 K_2$.
The $x$ and $y$-components of the first-order momentum equations can be manifested as
\begin{eqnarray}
&&\hspace*{-1.3cm}u_{1x}^{(1)}=-\frac{l_y v_p^2}{\Omega_{c1}(v_p^2 - \alpha\mu_1 l_z^2 K_1)} \frac{\partial\phi^{(1)}}{\partial\xi},
\label{eq:C4:32}\\
&&\hspace*{-1.3cm}u_{1y}^{(1)} =  \frac{l_x v_p^2}{\Omega_{c1}(v_p^2 - \alpha\mu_1 l_z^2 K_1)} \frac{\partial\phi^{(1)}}{\partial\xi},
\label{eq:C4:33}\\
&&\hspace*{-1.3cm}u_{2x}^{(1)}=-\frac{l_y v_p^2}{\Omega_{c1}(v_p^2 - \alpha\mu_1 l_z^2 K_2)} \frac{\partial\phi^{(1)}}{\partial\xi},
\label{eq:C4:34}\\
&&\hspace*{-1.3cm}u_{2y}^{(1)} =  \frac{l_x v_p^2}{\Omega_{c1}(v_p^2 - \alpha\mu_1 l_z^2 K_2)} \frac{\partial\phi^{(1)}}{\partial\xi}.
\label{eq:C4:35}\
\end{eqnarray}
Now, by taking the next higher-order terms, the equation of continuity, momentum equation, and Poisson's equation can be written as
\begin{eqnarray}
&&\hspace*{-1.3cm} \frac{ \partial n_1^{(1)}}{\partial \tau} - v_p  \frac{ \partial n_1^{(2)}}{\partial \xi}
+ l_x \frac{ \partial u_{1x}^{(1)}}{\partial \xi} + l_y \frac{ \partial u_{1y}^{(1)}}{\partial \xi}+ l_z \frac{ \partial u_{1z}^{(2)}}{\partial \xi}
\nonumber\\
&&\hspace*{0cm}+ l_z \frac{\partial}{\partial \xi} \Big(n_1^{(1)}  u_{1z}^{(1)} \Big) = 0,
\label{eq:C4:36}\\
&&\hspace*{-1.3cm} \frac{ \partial u_{1z}^{(1)}}{\partial \tau} - v_p \frac{ \partial u_{1z}^{(2)}}{\partial \xi} + l_z  u_{1z}^{(1)} \frac{\partial u_{1z}^{(1)}}{\partial \xi} + l_z \frac{\partial \phi^{(2)}}{\partial \xi}
\nonumber\\
&&\hspace*{-1cm}+ \alpha\mu_1 l_z K_1 \Biggr[ \frac{\partial n_1^{(2)}}{\partial \xi}
+ \frac{(\alpha-2)}{2}  \frac{\partial n_{1}^{(1)^2}}{\partial \xi} \Biggr] - \eta \frac{\partial^2 u_{1z}^{(1)}}{\partial \xi^2} = 0,
\label{eq:C4:37}\\
&&\hspace*{-1.3cm} \frac{ \partial n_2^{(1)}}{\partial \tau} - v_p  \frac{ \partial n_2^{(2)}}{\partial \xi}
+ l_x \frac{ \partial u_{2x}^{(1)}}{\partial \xi} + l_y \frac{ \partial u_{2y}^{(1)}}{\partial \xi} + l_z \frac{ \partial u_{2z}^{(2)}}{\partial \xi}
\nonumber\\
&&\hspace*{0cm}+ l_z \frac{\partial}{\partial \xi} \Big(n_2^{(1)}  u_{2z}^{(1)} \Big) = 0,
\label{eq:C4:38}\\
&&\hspace*{-1.3cm} \frac{ \partial u_{2z}^{(1)}}{\partial \tau} - v_p \frac{ \partial u_{2z}^{(2)}}{\partial \xi} +  l_z u_{2z}^{(1)}\frac{\partial u_{2z}^{(1)}}{\partial \xi} +\mu_2l_z \frac{\partial \phi^{(2)}}{\partial \xi}
\nonumber\\
&&\hspace*{-1cm}+ \alpha\mu_1l_z K_2 \Biggr[ \frac{\partial n_2^{(2)}}{\partial \xi}
+\frac{(\alpha-2)}{2}  \frac{\partial n_{2}^{(1)^2}}{\partial \xi} \Biggr] - \eta \frac{\partial^2 u_{2z}^{(1)}}{\partial \xi^2} = 0,
\label{eq:C4:39}\\
&&\hspace*{-1.3cm}\mu_4n_2^{(2)}+n_1^{(2)}=\sigma_1\phi^{(2)}+\sigma_2\phi^{(1)^2}.
\label{eq:C4:40}\
\end{eqnarray}
Finally, the next higher-order terms of Eqs. \eqref{eq:C4:10}-\eqref{eq:C4:13} and \eqref{eq:C4:16}, with the help of
Eqs. \eqref{eq:C4:26}-\eqref{eq:C4:40}, can provide the Burgers' equation as
\begin{eqnarray}
&&\hspace*{-1.3cm}\frac{\partial \Phi}{\partial \tau}+A\Phi\frac{\partial \Phi}{\partial \xi}=B\frac{\partial^2 \Phi}{\partial \xi^2},
\label{eq:C4:41}\
\end{eqnarray}
where $\Phi=\phi ^{(1)}$ for simplicity. In Eq. \eqref{eq:C4:41}, the nonlinear
coefficient $A$ and dissipative coefficient $B$ are, respectively, given by
\begin{eqnarray}
&&\hspace*{-1.3cm}A=P(Q+R-2\sigma_2),
\label{eq:C4:42}\\
&&\hspace*{-1.3cm}B=\frac{\eta}{2},
\label{eq:C4:43}\
\end{eqnarray}
where
\begin{eqnarray}
&&\hspace*{-1.3cm}P=\frac{(v_p^2-\alpha\mu_1l_z^2K_1)^2 (v_p^2-\alpha\mu_1l_z^2K_2)^2}{{2v_pl_z^2 [v_p^4(1+\mu_2\mu_4)+\alpha^2\mu_1^2l_z^4(K_2^2+\mu_2\mu_4K_1^2)-M]}},
\nonumber\\
&&\hspace*{-1.3cm}M=2\alpha\mu_1l_z^2v_p^2(K_2+K_1\mu_2\mu_4),
\nonumber\\
&&\hspace*{-1.3cm}Q=\frac{l_z^4\{3v_p^2+\mu_1l_z^2K_1\alpha(\alpha-2)\}}{(v_p^2-\alpha\mu_1l_z^2K_1)^3},
\nonumber\\
&&\hspace*{-1.3cm}R=\frac{\mu_2^2\mu_4l_z^4\{3v_p^2+\alpha \mu_1 \mu_2 l_z^2K_2(\alpha-2)\}}{(v_p^2-\alpha\mu_1l_z^2K_2)^3},
\nonumber\
\end{eqnarray}
Now, we look for stationary shock wave solution of this Burgers' equation  by
considering $\zeta =\xi-U_0\tau'$ and $\tau =\tau'$ (where $U_0$ is the speed of the shock waves in the reference frame).
These allow us to write the stationary shock wave solution as \cite{Hossen2017,Karpman1975,Hasegawa1975}
\begin{eqnarray}
&&\hspace*{-1.3cm}\Phi=\Phi_0\bigg[1-\tanh\bigg(\frac{\zeta}{\Delta}\bigg)\bigg],
\label{eq:C4:44}\
\end{eqnarray}
where the amplitude $\Phi_0$ and width $\Delta$ are given by
\begin{eqnarray}
&&\hspace*{-1.3cm}\Phi_0=\frac{U_0}{A},~~~~\mbox{and}~~~~\Delta=\frac{2B}{U_0}.
\label{eq:C4:45}\
\end{eqnarray}
It is clear from  Eqs. \eqref{eq:C4:44} and \eqref{eq:C4:45} that the IASHWs exist,
which are formed due to the balance between nonlinearity and dissipation,
because $B>0$ and the IASHWs with $\Phi>0$ ($\Phi<0$) exist if $A>0$ ($A<0$) because $U_0>0$.
\begin{figure}[t]
\centering
\includegraphics[width=80mm]{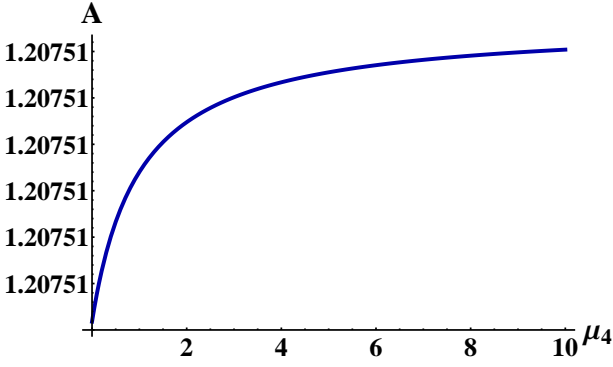}
\caption{Plot of the nonlinear coefficient $A$ vs $\mu_4$ when $\alpha=5/3$, $\delta=20^\circ$, $\eta=0.3$, $\gamma_e=4/3$, $Z_1=37$, $Z_2=6$, $n_{10}=10^{29}\mbox{cm}^{-3}$, $n_{20}=10^{30}\mbox{cm}^{-3}$, $n_{e_0}=10^{33}\mbox{cm}^{-3}$, and $v_p=v_{p+}$.}
\label{C4F1}
\end{figure}
\begin{figure}[t]
\centering
\includegraphics[width=80mm]{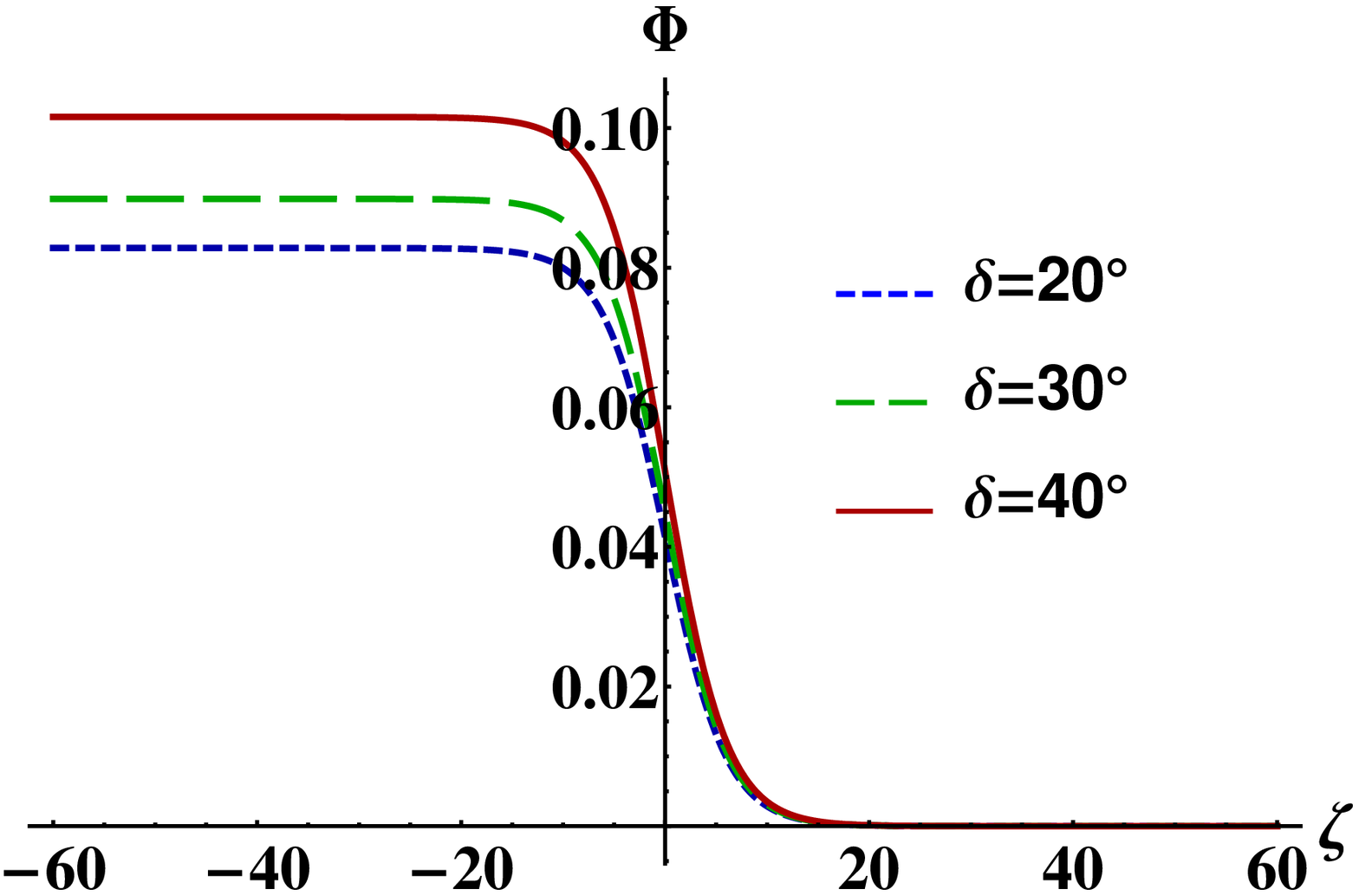}
\caption{Plot of $\Phi$ vs $\zeta$ for different values of $\delta$ when $\alpha=5/3$, $\delta=20^\circ$, $\eta=0.3$, $\gamma_e=4/3$, $Z_1=37$, $Z_2=6$,  $n_{10}=10^{29}\mbox{cm}^{-3}$, $n_{20}=10^{30}\mbox{cm}^{-3}$,  $U_0=0.05$, $n_{e_0}=10^{33}\mbox{cm}^{-3}$, and $v_p=v_{p+}$.}
\label{C4F2}
\end{figure}
\begin{figure}[t]
\centering
\includegraphics[width=80mm]{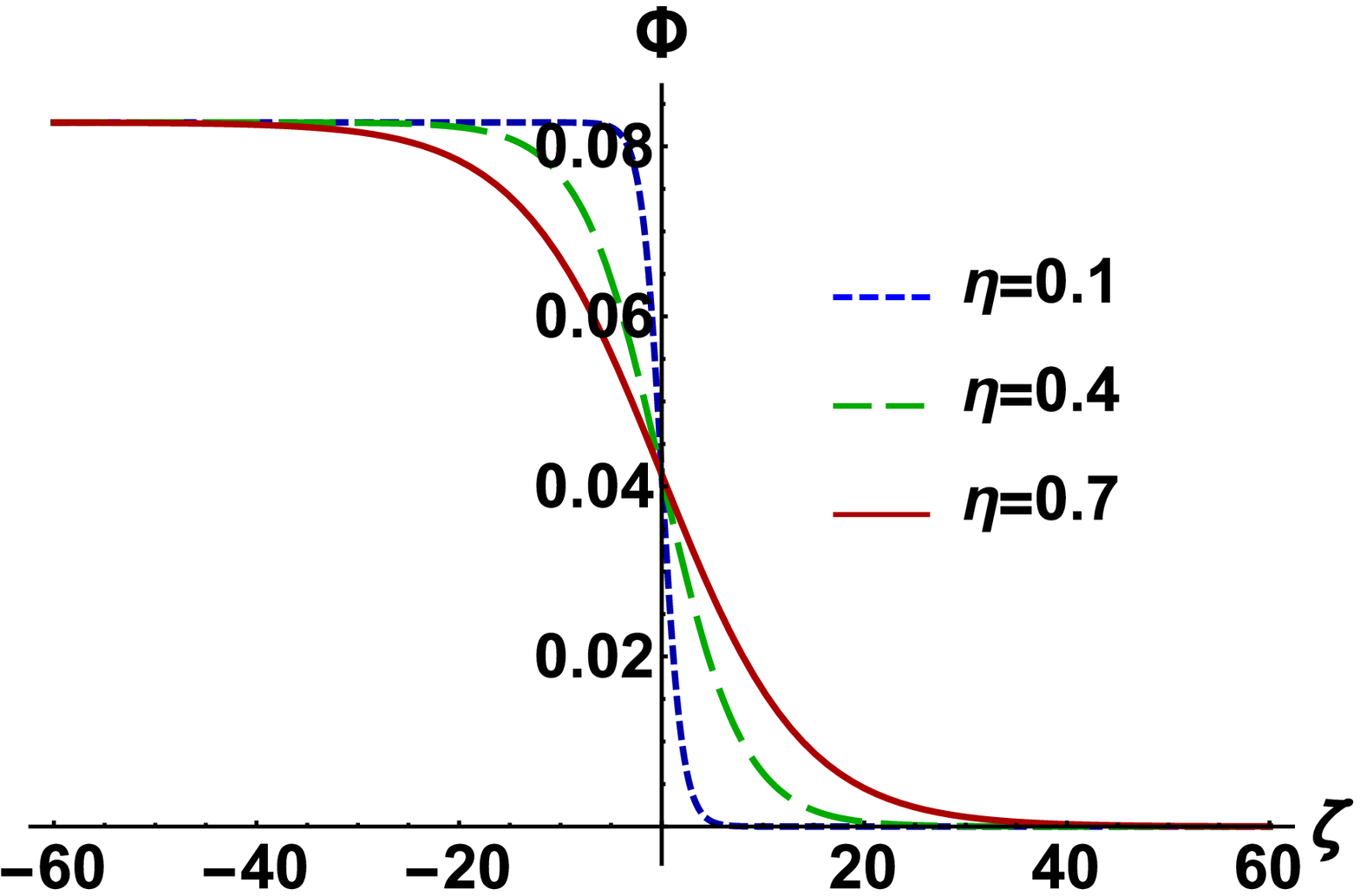}
\caption{Plot of $\Phi$ vs $\zeta$ for different values of $\eta$ when $\alpha=5/3$, $\delta=20^\circ$, $\eta=0.3$, $\gamma_e=4/3$, $Z_1=37$, $Z_2=6$,  $n_{10}=10^{29}\mbox{cm}^{-3}$, $n_{20}=10^{30}\mbox{cm}^{-3}$, $n_{e_0}=10^{33}\mbox{cm}^{-3}$, $U_0=0.05$, and $v_p=v_{p+}$.}
\label{C4F3}
\end{figure}
\begin{figure}[t]
\centering
\includegraphics[width=80mm]{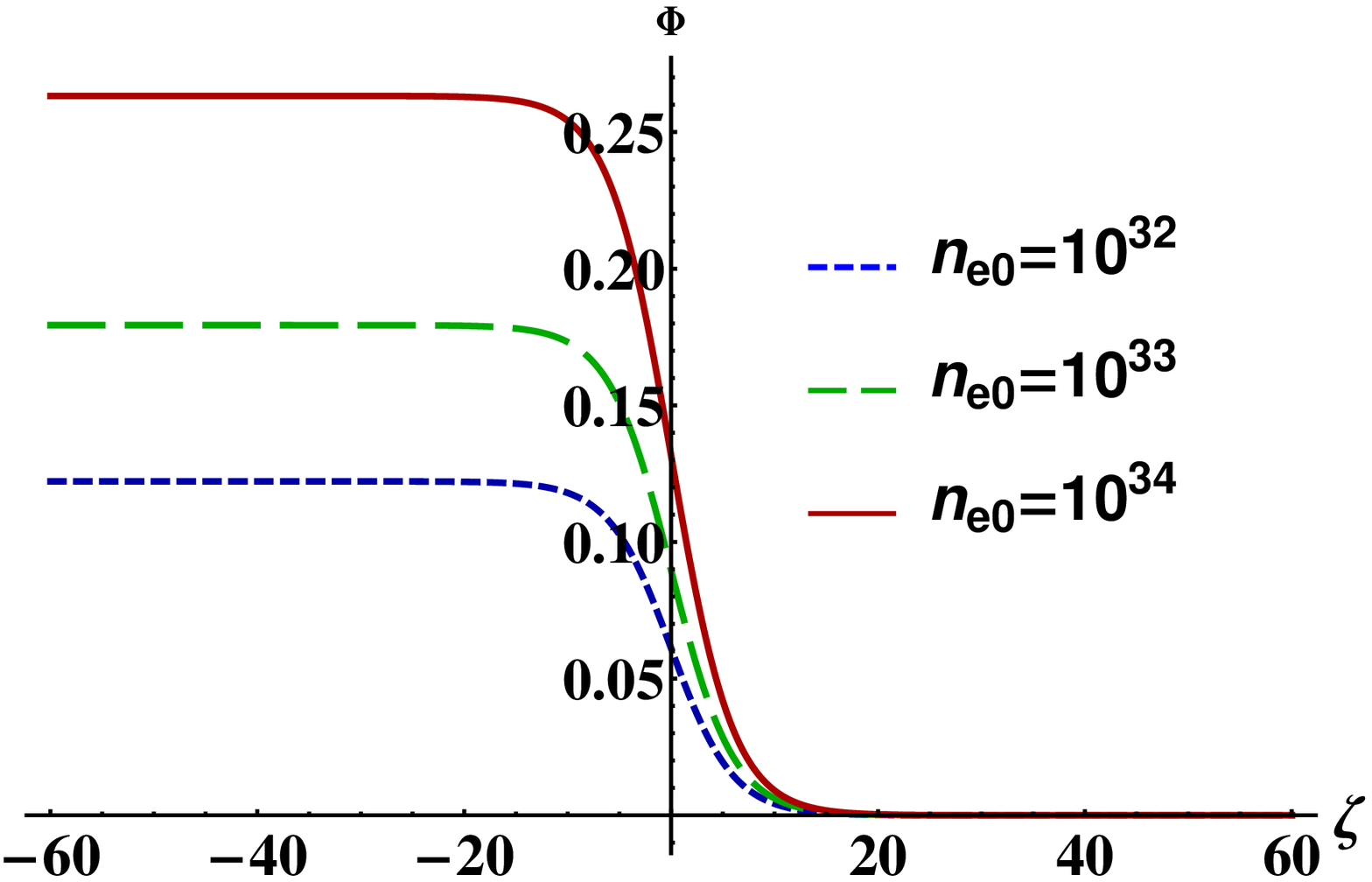}
\caption{Plot of $\Phi$ vs $\zeta$ for different values of $n_{e_0}$ when $\alpha=5/3$, $\delta=20^\circ$, $\eta=0.3$, $\gamma_e=4/3$, $Z_1=37$, $Z_2=6$, $n_{10}=10^{29}\mbox{cm}^{-3}$, $n_{20}=10^{30}\mbox{cm}^{-3}$,  $U_0=0.05$, and $v_p=v_{p+}$.}
\label{C4F4}
\end{figure}
\begin{figure}[t]
\centering
\includegraphics[width=80mm]{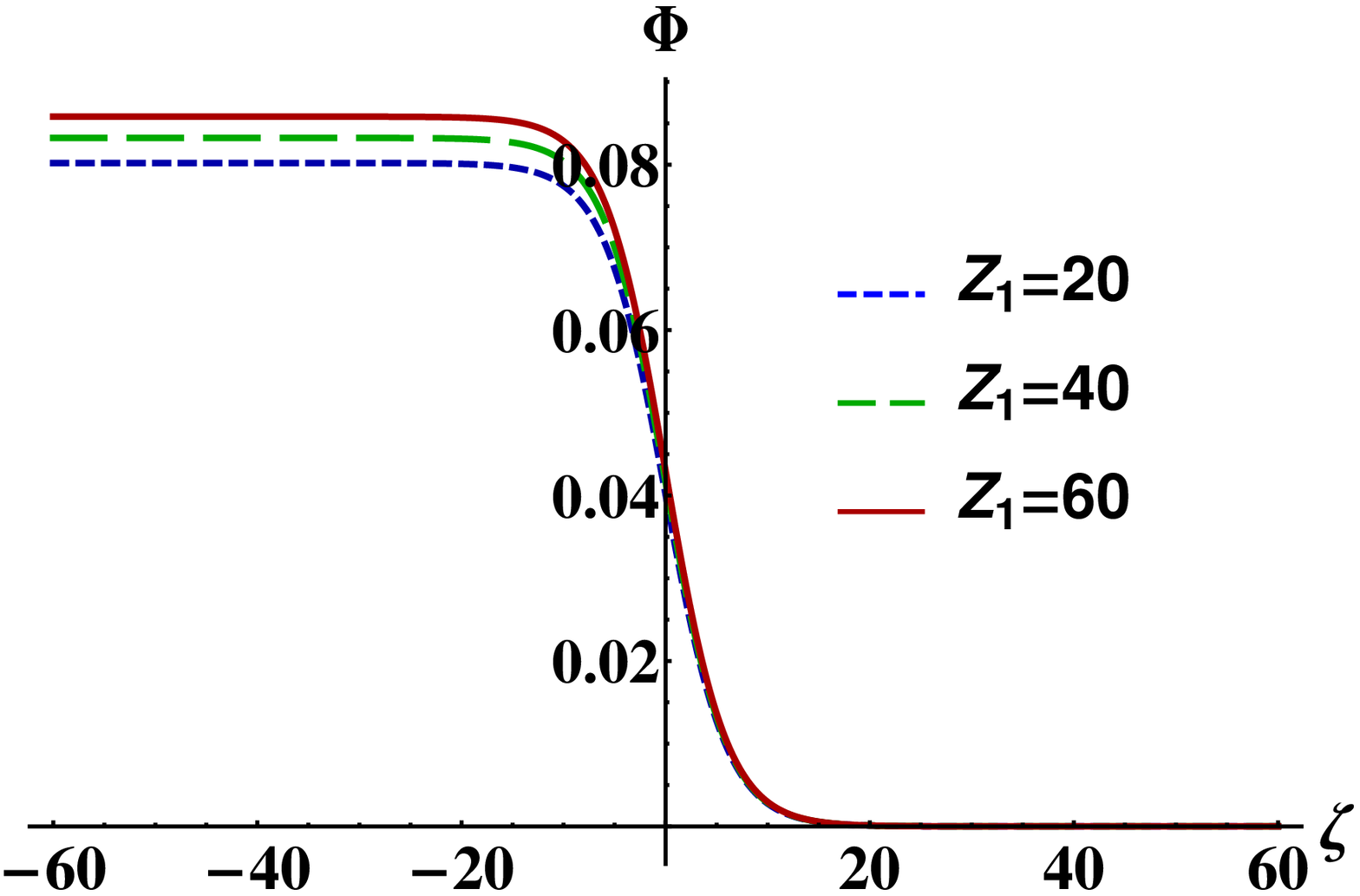}
\caption{Plot of $\Phi$ vs $\zeta$ for different values of $Z_1$ when $\alpha=5/3$, $\delta=20^\circ$, $\eta=0.3$, $\gamma_e=4/3$, $Z_2=6$,  $n_{10}=10^{29}\mbox{cm}^{-3}$, $n_{20}=10^{30}\mbox{cm}^{-3}$, $n_{e_0}=10^{33}\mbox{cm}^{-3}$, $U_0=0.05$, and $v_p=v_{p+}$.}
\label{C4F5}
\end{figure}
\begin{figure}[t]
\centering
\includegraphics[width=80mm]{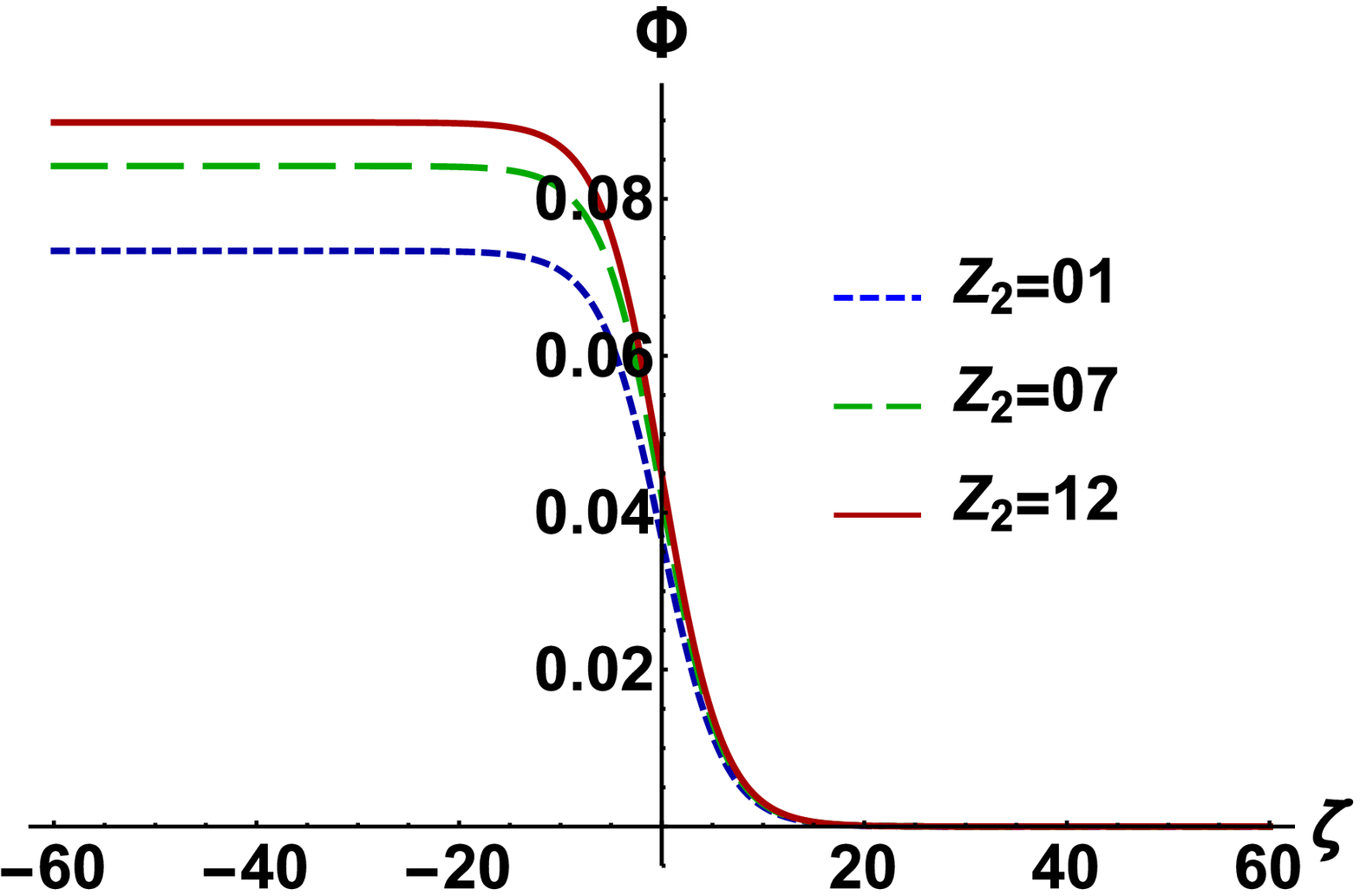}
\caption{Plot of $\Phi$ vs $\zeta$ for different values of $Z_2$ when $\alpha=5/3$, $\delta=20^\circ$, $\eta=0.3$, $\gamma_e=4/3$, $Z_1=37$,  $n_{10}=10^{29}\mbox{cm}^{-3}$, $n_{20}=10^{30}\mbox{cm}^{-3}$, $n_{e_0}=10^{33}\mbox{cm}^{-3}$, $U_0=0.05$, and $v_p=v_{p+}$.}
\label{C4F6}
\end{figure}
\begin{figure}[t]
\centering
\includegraphics[width=80mm]{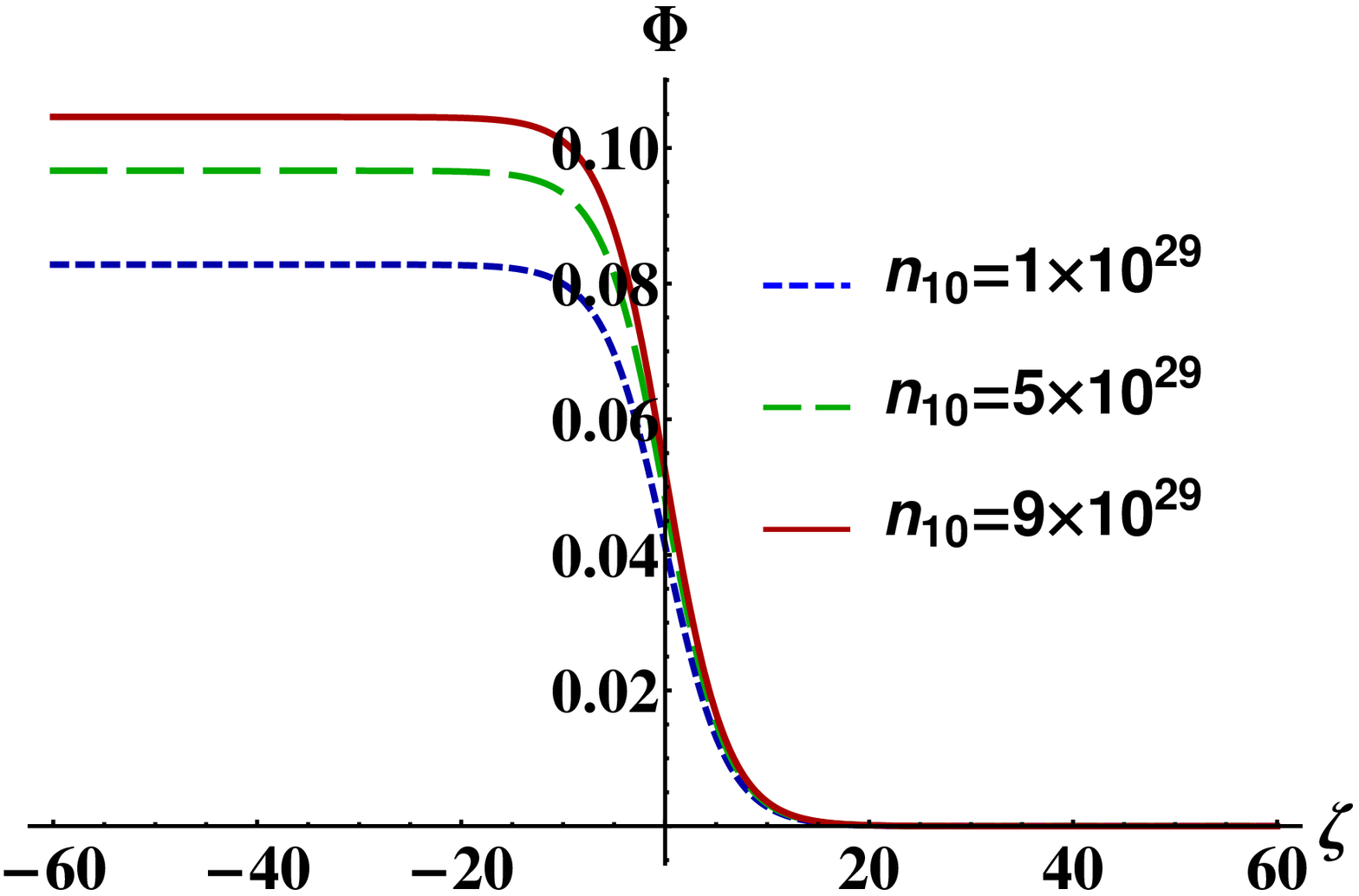}
\caption{Plot of $\Phi$ vs $\zeta$ for different values of $n_{10}$ when $\alpha=5/3$, $\delta=20^\circ$, $\eta=0.3$, $\gamma_e=4/3$, $Z_1=37$, $Z_2=6$,  $n_{10}=10^{29}\mbox{cm}^{-3}$, $n_{20}=10^{30}\mbox{cm}^{-3}$, $n_{e_0}=10^{33}\mbox{cm}^{-3}$, $U_0=0.05$, and $v_p=v_{p+}$.}
\label{C4F7}
\end{figure}
\begin{figure}[t]
\centering
\includegraphics[width=80mm]{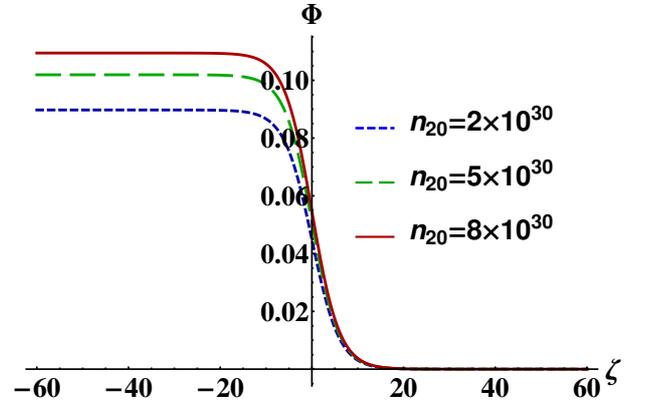}
\caption{Plot of $\Phi$ vs $\zeta$ for different values of $n_{20}$ when $\alpha=5/3$, $\delta=20^\circ$, $\eta=0.3$, $\gamma_e=4/3$, $Z_1=37$, $Z_2=6$,  $n_{10}=10^{29}\mbox{cm}^{-3}$, $n_{e_0}=10^{33}\mbox{cm}^{-3}$,  $U_0=0.05$, and $v_p=v_{p+}$.}
\label{C4F8}
\end{figure}
\section{Results and Discussions}
\label{2sec:Results and Discussions}
Our present investigation is valid for white dwarfs and neutron stars in which both non-relativistic
positively charged heavy ions (e.g., ${}^{56}_{26}{\mbox{Fe}}$ \cite{Vanderburg2015},
${}^{85}_{37}{\mbox{Rb}}$ \cite{Witze2014}, ${}^{96}_{42}{\mbox{Mo}}$ \cite{Witze2014}), and
light ions (e.g., ${}^{1}_{1}{\mbox{H}}$ \cite{Fletcher2006,Killian2006}, ${}^{4}_{2}{\mbox{He}}$ \cite{Flower1994},
${}^{12}_{6}{\mbox{C}}$ \cite{Koester1990,Koester2002}), and ultra-relativistically degenerate electrons are exist.
For numerical analysis, we have considered $Z_1=20\thicksim60$, $Z_2=1\thicksim12$,  $n_{10}=1\times10^{29}\mbox{cm}^{-3}\thicksim9\times10^{29}\mbox{cm}^{-3}$, $n_{20}=2\times10^{30}\mbox{cm}^{-3}\thicksim8\times10^{30}\mbox{cm}^{-3}$, and $n_{e_0}=10^{32}\mbox{cm}^{-3}\thicksim10^{34}\mbox{cm}^{-3}$.
The IASHW is governed by the Burgers' equation \eqref{eq:C4:41}, and the positive (negative) shock potential can exist corresponding to the limit of $A>0$ ($A<0$). The variation of $A$ with $\mu_4$ can be seen from Fig. \ref{C4F1}, and it is clear from this figure that our plasma model supports only
positive shock potential under consideration of both non-relativistic
positively charged heavy and light ions (i.e., $\alpha=5/3$), and ultra-relativistically degenerate electrons  (i.e., $\gamma_e=4/3$).

The parameter $\delta$ reveals the angle between the direction of the wave propagation and the direction of the external
magnetic field, and the effects of $\delta$ on the formation of IASHWs can be seen in Fig. \ref{C4F2}. When the
oblique angle ($\delta$) increases, the
magnetic effect becomes more significant, and therefore the amplitude of the shock wave increases, and this result agrees with the
result of Hossen \textit{et al.} \cite{Hossen2017}.

Figure \ref{C4F3} illustrates the effects of the non-relativistic heavy and light ion's kinematic viscosity
on the positive potential (i.e., $\Phi>0$) under consideration of $A>0$. It is
really interesting that the steepness of the shock profile decreases with an increase in the value of the
non-relativistic heavy and light ion's kinematic viscosity but the amplitude of shock profile is not
affected by the kinematic viscosity ions, and this result
agrees with the previous work of Abdelwahed \textit{et al.} \cite{Abdelwahed2016}.

The variation of IASHWs with electron number density ($n_{e0}$) under
consideration of both non-relativistic positively charged heavy and light ions (i.e., $\alpha=5/3$),
and ultra-relativistically degenerate electrons  (i.e., $\gamma_e=4/3$) can be observed in  Fig. \ref{C4F4}.
It is clear from this figure that as we increase the electron number density, the amplitude of the IASHWs
associated with $\Phi>0$ (i.e., $A>0$) increases. So, the ultra-relativistic electrons enhance the amplitude of the
IASHWs in a magnetized DQPS having non-relativistic positively charged heavy and light ions, and ultra-relativistically
degenerate electrons.

The effects of the charge state of non-relativistic heavy and light ions species on the formation of IASHWs in a magnetized
DQPS can be seen in Fig. \ref{C4F5}  and \ref{C4F6}, respectively. It is obvious from these figures that the charge state of
both non-relativistic heavy and light ion species enhances the amplitude of IASHWs associated with $\Phi>0$ (i.e., $A>0$) under
consideration of $\alpha=5/3$ and $\gamma_e=4/3$. Physically, both non-relativistic heavy and light ion species, due to both are positively
charged,  play same role in the dynamics of magnetized DQPS as well as the configuration of IASHWs. Similarly, the number density of
the non-relativistic heavy and light ion species can play significant role in the formation of IASHWs. It is clear form Figs.  \ref{C4F7}  and \ref{C4F8} that the amplitude of the IASHWs associated with $\Phi>0$ (i.e., $A>0$) and under
consideration of $\alpha=5/3$ and $\gamma_e=4/3$ increases with the number density of both non-relativistic heavy and light ion species.
\section{Conclusion}
\label{2sec:Conclusion}
We have investigated the fundamental characteristics of IASHWs in a magnetized DQPS having
inertial non-relativistic positively charged heavy and light ions, inertialess
ultra-relativistically degenerate electrons. The reductive perturbation method \cite{C1}has been
employed to derive Burgers' equation. The results which have been found from present study
can be pinpointed as follows:
\begin{itemize}
  \item The plasma model supports only positive shock potential under consideration of both non-relativistic
        positively charged heavy and light ions (i.e., $\alpha=5/3$), and ultra-relativistically degenerate electrons  (i.e., $\gamma_e=4/3$).
  \item The greater number density of ultra-relativistic electrons enhances the amplitude of the IASHWs.
  \item The increasing charge state and number density of the non-relativistic heavy and light ion species enhance
  the amplitude of the IASHWs associated with $\Phi>0$ (i.e., $A>0$).
  \item The steepness of the shock profile is decreased with the kinematic viscosity ($\eta$) of ions.
  \item The amplitude of the shock profile is found to increase as the oblique angle increases.
\end{itemize}
It may be noted here that the self-gravitational effects of the DQPS are really important to include in the
governing equations but beyond the scope of our present work. However, We are optimistic that the outcomes
from our present investigation will be useful to understand the propagation of IASHWs in white dwarfs and neutron stars in which
the non-relativistic positively charged heavy and light ions, and ultra-relativistically degenerate electrons are exist.
\section*{Acknowledgments}
Authors would like to acknowledge ``UGC research project 2018-2019'' for their financial supports
to complete this work.

\end{document}